\documentclass[a4paper]{article}

\usepackage{a4wide}
\usepackage{mathtools}
\usepackage[utf8]{inputenc}
\usepackage{natbib}
\usepackage{graphicx}
\usepackage{accents}
\usepackage{comment}
\usepackage{amsmath,amssymb}
\usepackage{multirow}
\usepackage{booktabs}
\usepackage{algpseudocode}
\usepackage{algorithm}

\usepackage{rotating}
\usepackage{ dsfont }
\usepackage{wrapfig}
\usepackage{lscape}
\usepackage{epstopdf}
\usepackage{xcolor}
\usepackage{authblk}

\DeclareMathOperator{\EX}{\mathbb{E}}% expected value

\DeclareMathOperator{\argmin}{arg\,min}

\title{\vspace{-2cm}\LARGE\centering\textbf{The \textsc{CDF} penalty: \\ sparse and quasi unbiased estimation in regression models}}
\author[1,*]{Daniele Cuntrera}
\author[1]{Luigi Augugliaro}
\author[1]{Vito M.R. Muggeo}
\affil[1]{University of Palermo, Department Economics, Business and Statistics, Palermo, Italy.}
\affil[*]{Corresponding author: daniele.cuntrera@unipa.it}

\begin{document}

\maketitle

\begin{abstract}
In high-dimensional regression modelling, the number of candidate covariates to be included in the predictor is quite large, and variable selection is crucial. In this work, we propose a new penalty able to guarantee both sparse variable selection, i.e. exactly zero regression coefficient estimates, and quasi-unbiasedness for the coefficients of 'selected' variables in high dimensional regression models. Simulation results suggest that our proposal performs no worse than its competitors while always ensuring that the solution is unique.
\end{abstract}


\section{Introduction}

The 21st century was characterized by technological development that led to the spasmodic collection of massive amounts of data. Technological evolution and advances in information technology over the last two decades have led to the arrangement of a new way of thinking about storing large amounts of data, leading to the advent of high and ultra-high-dimensional data. This gives rise to the need to formalize new statistical techniques to select a subset of covariates helpful in explaining the phenomenon under study. This problem spans various application areas \citep{donoho2000high}: it typically arises in the field of genetics but is found in a broader sense in the biotechnology sector or finance. There is some context in which the number of predictors is much larger than the observation: this involves problems like collinearity problems \citep{fan2008sure} or noise accumulation \citep{fan2008high, hall2008theoretical}.   

Until the advent of high and ultra-high dimensionality, statistical models were based on three foundations, i.e. accuracy, interpretability, and not too much computational complexity \citep{fan2010selective}. With the development of sciences and the constant growth of data availability, classical statistical methods do not fit new challenges: using these involves sacrificing at least one of the three pillars.

However, using methods that estimate a ``sparse model'' (i.e. a model that reduces the dimension of the estimated parameters) improves all three aforementioned foundations due to the lower dimensionality that enhances the ease of interpretation and the accuracy of non-null coefficients. Different approaches exist to perform variable selection and produce sparse models, like test-based (e.g. \cite{breaux1967stepwise, hendry1987recent}) or screening-based approaches (e.g. \cite{fan2008sure}). One of the drawbacks of these methods is that they alternate the phases of selection and estimation of coefficients, with significant consequences on computational complexity; a third approach that can be used is the penalized-based: it is performed considering a penalization on the coefficients of a regression model, just modifying the objective function to minimize. If a penalty is well defined, the model can perform variable selection, and the selection and estimation of parameters are simultaneous. In an overparametrized model, regularisation reduces the model size simply by using data, avoiding concerns for overfitting and improving the model's accuracy and interpretability. The goal of this paper is to introduce a new penalty function, or better a family of penalties indexed by a shape parameter which enhances flexibility for practical applications. Our proposal lies somewhere between convex and non-convex penalties and it recovers the well-known LASSO as the limiting case. The rest of the paper is organized as follows: in section 2 we discuss the methodological proposal, and in section 3 we present evidence from some simulation experiments.

\subsection{Penalized GLM framework}

Let $(X, Y)$ be a collection of random variables, where $X$ is a $p$-dimensional vector of predictor variables while the scalar $Y$ is the dependent variable. 

As usual, we assume a random sample of size $n$ drawn from the distribution $(X, Y)$, where the density of the conditional distribution of $ (Y | X = x)$ comes from a one-parameter exponential family distribution expressed as  $f(y;\theta) = \exp{ \left( \frac{y \theta - b(\theta)}{a(\phi)} + c(y, \phi) \right) }$ where $a(\cdot)$, $b(\cdot)$ and $c(\cdot)$ are known functions, and $\theta$ is the natural parameter. To study the effect of the explanatory variables on the response variable, it is imposed that the parameter $\theta$ is a function of $X$, i.e.

$$ f(y | x ;\theta) = \exp{ \left( \frac{y \theta(x) - b(\theta(x))}{a(\phi)} + c(y, \phi) \right) } \quad \theta \in \mathds{R} , \quad \phi \in \mathds{R}^+$$

The goal of mean regression modelling is to model the expectation of Y$|x$

$$ \EX(Y|X = x) = b'(\theta(x)) = \mu(x). $$

The conditional expected value is linked to the covariates through the link function transformation $g(\mu(x_i)) = x^T_i\beta$, referred to as the linear predictor. Usually, regression analysis aims at estimating $\beta$ that expresses the effect of the covariates on $\mu$. Estimation is carried out by maximizing the likelihood, but when the number of covariates is large or $p > n$, a penalty is added to shrink the parameter estimates.

The penalized log-likelihood is defined as usual

\begin{equation} \label{eq:penlik}
    \ell_\lambda(\beta) = \ell(\beta) - \sum_{j=1}^p p_{\lambda} \left( |\beta_j| \right), 
\end{equation}
see \cite{o1986automatic}, \cite{gu2002penalized}, and  \cite{fan2001variable} among others. The term $p_{\lambda} \left( |\beta_j| \right)$ is the penalty function which allows to select and estimate coefficients simultaneously. The penalized likelihood is indexed by the tuning parameter $\lambda$ that influences the complexity of the model. If $\lambda=0$, an unpenalized model is fitted, while for large values of $\lambda$, the model becomes more sparse until no coefficient is estimated. Typically, the selection of $\lambda$ is made by optimizing a measurement of prediction error like AIC or BIC \citep{lv2014model, chen2008extended} or via cross-validation.

\subsection{Related works and their limitations}
%Convesse / non convesse

The literature on penalty functions is now boundless, and to conduct an exhaustive review of them would require dedicated work. See, for example, \cite{fan2010selective} for broad perspectives on high-dimensional statistical problems.

Roughly speaking, an estimator produced by a proper penalty function should enjoy the three properties stated by \cite{fan2001variable}:

\begin{enumerate}
    \item sparsity: the estimator can reduce the number of parameters, thus setting 0 the noise-source coefficients. A sufficient condition is that $\argmin_{\beta} \left[ |\beta| + p'_{\lambda}(|\beta|) \right]$ is positive;
    \item continuity: the estimator is continuous in the data; in other words, a slight change in the data should not result in a significant change in the estimates. A necessary and sufficient condition for the penalty to be continuous is that $\argmin_{\beta} \left[ |\beta| + p'_{\lambda}(|\beta|) \right] = 0$;
    \item unbiasedness: the estimator is nearly unbiased for large coefficients; a sufficient condition is that $\lim_{\beta \to \infty} p'_{\lambda}(|\beta|) = 0$.
\end{enumerate}

%Additionally, from a computational perspective, penalty functions should be chosen so that the resulting optimization problem is straightforward to solve. 
In literature some options are available, see Table \ref{tab:pen}.

%Focussing only on the penalties able to select variables, it is possible to divide the penalizations into two macro-categories: the former concerns approach based on the use of convex penalty functions, e.g. LASSO \citep{tibshirani1996regression}, adaptive LASSO \citep{zou2006adaptive}, Non-negative Garrotte \citep{breiman1995better}; the latter involves non-convex penalty functions, e.g. Bridge estimator \citep{frank1993statistical}, SCAD \citep{fan2001variable}, MCP \citep{zhang2010nearly}, log-penalty \citep{friedman2012fast}. Table \ref{tab:pen} lists some examples of such penalties. 

\begin{table}[!hb] \small 
  \
  \caption{\label{tab:pen} Some $L_1$-type penalties allowing variable selection in regression models.}
  \hfill
  \begin{center}
    \begin{tabular}{llll}\hline 
    \multicolumn{1}{l}{Name and reference} &       & \multicolumn{1}{c}{Penalty} & \multicolumn{1}{c}{} \\
    \midrule
    1. Hard thresholding * & & $p_{\lambda} \left( |\beta| \right) = \lambda I(|\beta| >  0)$  &\\
    $\quad$ \cite{antoniadis1997wavelets} & &   \\
    & &  & \\

    2. Bridge ** & & $p_{\lambda} \left( |\beta| \right) = \lambda |\beta|^{\gamma}$ & $\gamma \in (0,1)$  \\
    $\quad$ \cite{frank1993statistical} & & &  \\
    & & &  \\
    
    3. LASSO * & & $p_{\lambda} \left( |\beta| \right) = \lambda |\beta|$ &\\
    $\quad$ \cite{tibshirani1996regression} & & &  \\
    & & &  \\
    
    4. Adaptive LASSO * & & $p_{\lambda} \left( |\beta| \right) = \lambda w |\beta|$&\\
    $\quad$ \cite{zou2006adaptive} & & &  \\
    & & &  \\
    
    5. SCAD ** & & $p'_{\lambda} \left( |\beta| \right) =$   & \\
    $\quad$ \cite{fan2001variable} & & $\quad =\lambda \left\{ I_{|\beta| \le \lambda} + \frac{(\gamma \lambda - |\beta|)_+}{(\gamma - 1)\lambda} I_{|\beta| > \lambda} \right\}$ & $\gamma > 2$    \\
    & & &  \\
    
    6. MCP **& &  $p'_{\lambda} \left( |\beta| \right) = \lambda \left( 1 - \frac{|\beta|}{\gamma \lambda} \right) I_{|\beta| \le \lambda}$  & $\gamma > 1$ \\
    $\quad$ \cite{zhang2010nearly} & & &   \\
    & & &  \\
    
    7. Log-penalty **& & $p_{\lambda} \left( |\beta| \right) = \frac{\lambda}{\log(\gamma + 1)} \log(\gamma |\beta| + 1)$   & $\gamma > 0$\\
    $\quad$ \cite{friedman2012fast} & & &  \\
    & &  & \\
\bottomrule
\multicolumn{3}{l}{For SCAD and MCP the additional parameters are $\gamma_{SCAD} = 3.7$ and $\gamma_{MCP} = 3$.}
\\ \multicolumn{3}{l}{*Convex penalty; **Non-convex penalty}
\end{tabular}%
  \end{center}
    \end{table}%
%\begin{figure}[ht]
%    \centering
%    \includegraphics[width = \textwidth]{Derivate.pdf}
%    \caption{Derivative of SCAD (a), MCP (b) and Bridge (c).}
%    \label{fig:derv}
%\end{figure}
The LASSO is the most popular among the convex penalties. It enjoys excellent computational properties, and stable algorithms can be easily implemented without requiring complex optimizations. An example is the \textit{Least Angle Regression} (LARS) algorithm \citep{efron2004least}, providing the whole solution path as a function of the tuning parameter $\lambda$. Another popular algorithm is the \textit{Coordinate Descent} \citep{friedman2010regularization}, which sequentially updates coefficients cyclically and separately, maintaining the other coefficients as constants. The LASSO penalty enjoys the properties of sparsity and continuity but not unbiasedness because $\lim_{\beta \to \infty} p'_{LASSO, \lambda}(|\beta|) = \lambda$. Several authors have carefully examined the LASSO model consistency \citep{fu2000asymptotics, zou2006adaptive, zhao2006model}: broadly speaking the LASSO recovers the true sparse subset of variables only if certain conditions on the covariance among covariates (both true than noisy) are verified. In particular, \cite{zou2006adaptive} demonstrated that the inequality below must hold for LASSO to be consistent in variable selection

$$ | cov(X_\mathrm{noisy}, X_\mathrm{true}) cov(X_\mathrm{true})^{-1} s_\mathrm{true}| \le 1, $$
where inequality is intended component-wise and $s_\mathrm{true}$ is the vector of signs of the true parameter values.

The non-convex penalties lead to estimators with low bias, especially when the true coefficients are large. SCAD and MCP are probably the most widespread, as they enjoy the oracle properties, i.e. the estimator is asymptotically equivalent to that obtained by removing all noisy variables beforehand \citep{fan2001variable, fan2004nonconcave}. However, optimization with both the SCAD and MCP penalty, is not straightforward: approximations such as Local Quadratic Approximation (LQA) \citep{fan2001variable} and Local Linear Approximation (LLA) \citep{zou2008one} have been proposed to apply algorithms for weighted convex penalties. Other contributions include the PLUS algorithm by \cite{zhang2010nearly}, \cite{mazumder2011sparsenet} which propose a new version of Coordinate Descent, and \cite{breheny2011coordinate} which still rely on coordinate descent algorithms. Despite their excellent properties, non-convex penalties are not as widespread as the convex ones. Actually, non-convex penalties suffer from the non-uniqueness of the solution, in that the above-mentioned methods can stop at local, rather than global, optima \citep{kim2008smoothly}. Consequently, the solution path is not unique and could not contain the oracle estimator. This issue arises because the convexity of the likelihood does not predominate the concavity of the penalty part, leading to a non-convex penalized objective. For instance, \cite{breheny2011coordinate} highlight the importance of convexity diagnostics in identifying areas of the parameter space where the objective function is locally convex.

\section{The proposed penalty}

We introduce a new class of penalty functions $ p_{\lambda} \left( |\beta_j| \right) \propto \lambda F(|\beta_j|)$, where $F(\cdot)$ is the cumulative distribution function of a generic random variable. In this work, we propose the \textsc{CDF} penalty, a penalty function proportional to the cumulative density function of the standard Normal distribution,

%For the probability distribution to be correctly chosen, it must meet two conditions:

%begin{itemize}
%    \item the distribution function must be of a continuous and invertible random variable;
%    \item the distribution function must be differentiable in $\mathds{R}$ at least to the first order.
%\end{itemize}

%The first condition ensures that a coefficient can be penalized for any estimated value; the second allows the penalized model to be estimated without computational problems.

\begin{equation} \label{eq:proposal}
\ p_{\lambda}(|\beta_j|) = \lambda \sqrt{2\pi} \nu \Phi\left( \frac{|\beta_j|}{\nu} \right).
\end{equation}

It is worth stressing that the choice of standard Normal distribution is not due to some assumption about the distribution of coefficients but only to the penalty shape. The absolute value of the parameter ensures the singularity at the origin so that the penalty can perform the variable selection as those in Table \ref{tab:pen}. The \textsc{CDF} penalty lies between the convex and a non-convex penalty and therefore it is expected to share some of the good features of SCAD and MCP penalties, nearly unbiasedness while maintaining the stability of the LASSO.

\begin{figure*}[!hb] \label{fig:penalties}
    \centering
    \includegraphics[width=0.9\textwidth]{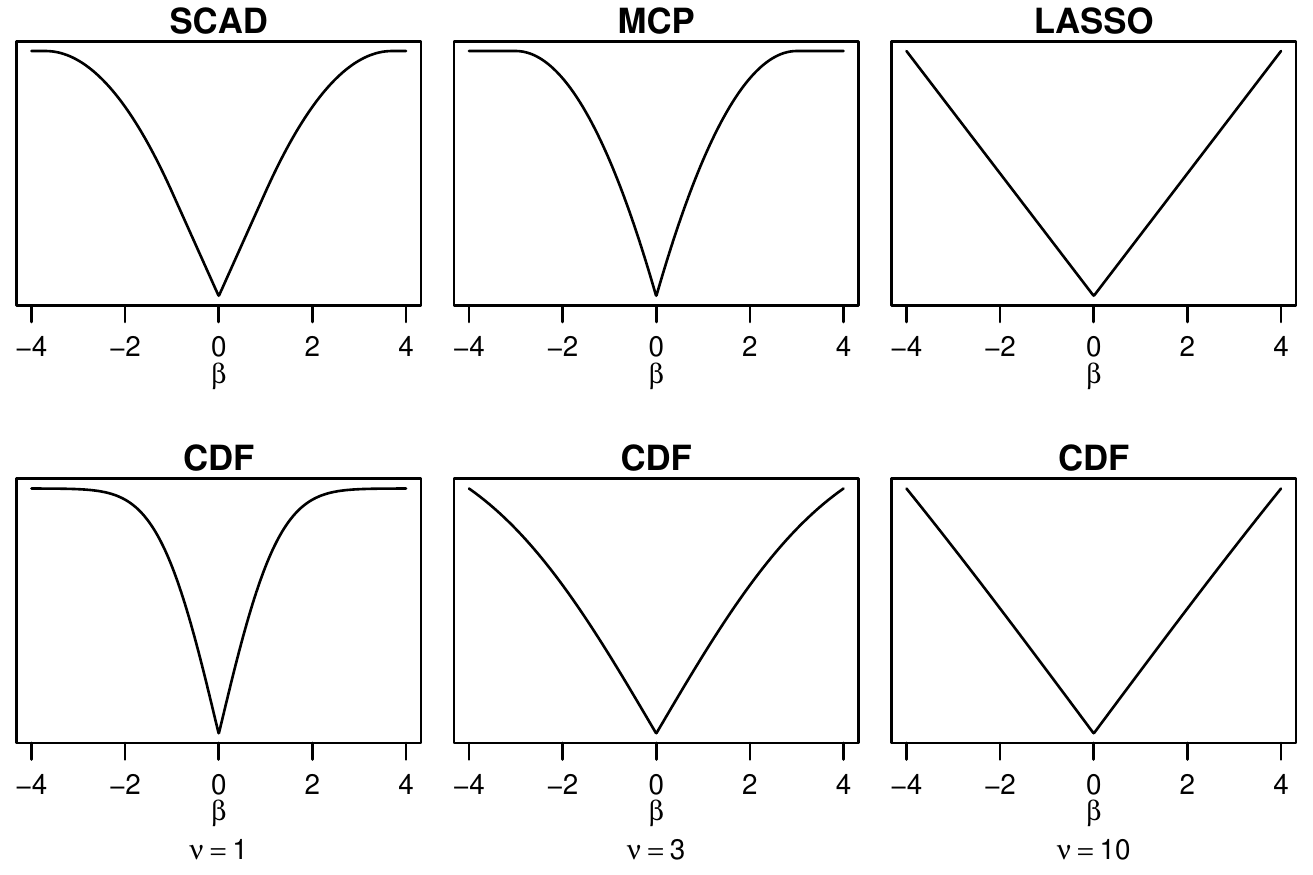}
    \caption{The shape of the four discussed penalties. Top: SCAD ($\lambda=1, \gamma = 3.7$), MCP ($\lambda=1, \gamma = 3$) and LASSO. Bottom: \textsc{CDF} for different values of $\nu$ (see text).}
\end{figure*}
Figure \ref{fig:penalties} shows the SCAD, MCP, LASSO and the \textsc{CDF} penalties at three different values of $\nu$ which affects the ``degree of non-convexity'' of the penalty: the larger $\nu$, the negligible the non-convexity.

%  For SCAD and MCP $\gamma$ is set respectively to 3.7 and 3, as suggested in literature \cite{fan2001variable, zhang2010nearly}. It is worth noting that the \textsc{CDF} penalty looks like SCAD and MCP if $\nu$ is small, while as $\nu$ gets larger the \textsc{CDF} shape becomes similar to LASSO.

\begin{figure}[!ht]
    \centering
    \includegraphics[width = .6\textwidth]{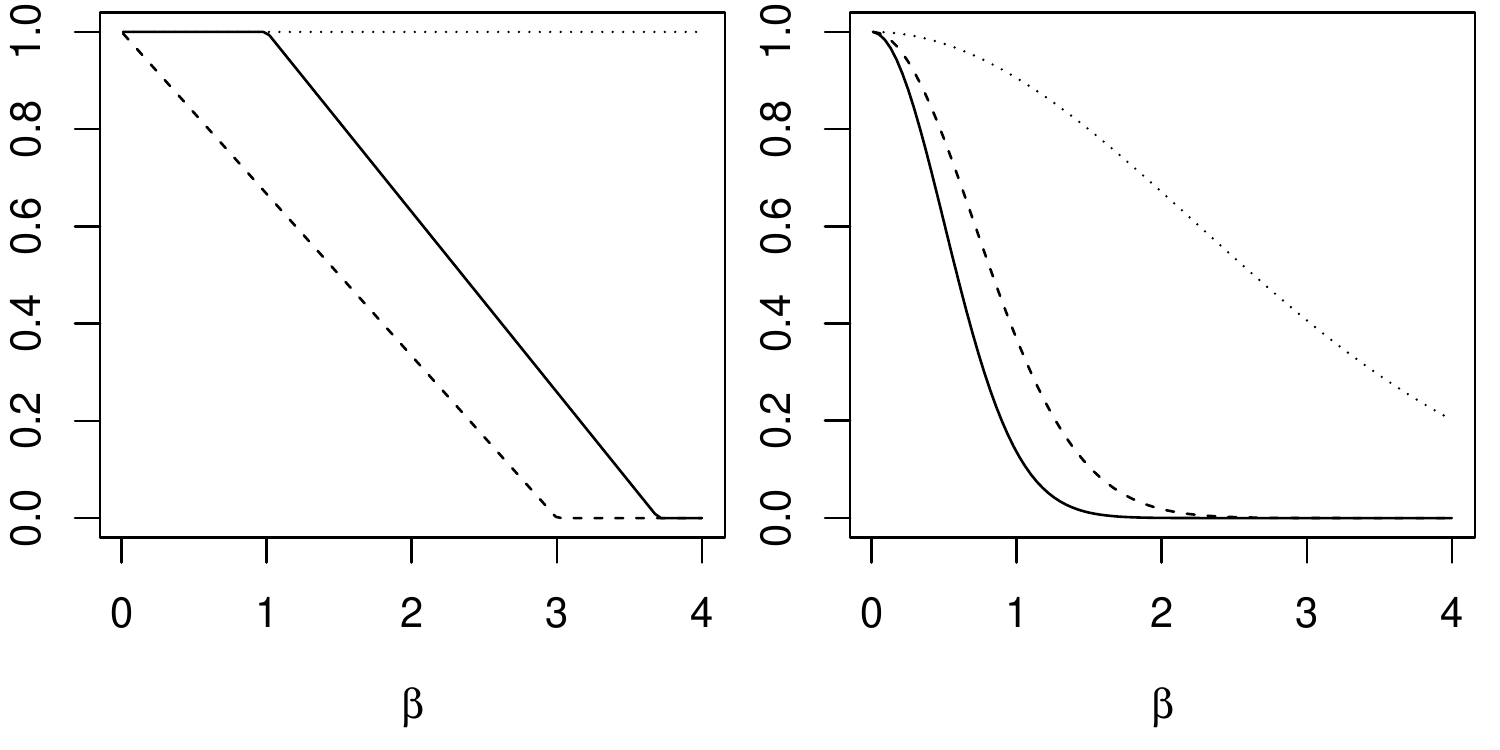}
    \caption{Left panel: derivative of SCAD ($\gamma$ = 3.7, continuous line), MCP ($\gamma$ = 3, dashed line) and LASSO (dotted line). Right panel: derivative of \textsc{CDF} using $\nu$ 0.5 (continuous line), 1 (dashed line) and 10 (dotted line).}
    \label{fig:derv}
\end{figure}

It is easy to see the \cite{fan2001variable} conditions discussed in section 2.2 are fulfilled, and moreover, the marked non-convexity at small $\nu$ ensures unbiasedness of the non-zero estimates as in SCAD and MCP. However the proposed CDF penalty (\ref{eq:proposal}) offers some additional advantages: \\
i) it is multiplicative respect to $\lambda$, namely $p_{\lambda}(\cdot)=\lambda \ p(\cdot)$ making the penalty shape independent of $\lambda$ unlike SCAD and MCP penalties whose shape itself does depend on $\lambda$ (and also on the additional $\gamma$, see Table 1); \\
ii) it is infinitely differentiable (see Figure \ref{fig:derv}); \\
iii) it is flexible, i.e. the amount of convexity can be tuned by the additional parameter $\nu$ (encompassing the LASSO as limiting case), and more importantly\\
iv) it always guarantees the uniqueness of the solution. 

The last point deserves some comments. The choice of $\nu$ is crucial as it affects both computational and inferential aspects. As seen in previous sections, $\nu$ affects the ``degree of non-convexity'' of the penalty and determines the amount of bias of the non-null estimates. However, it turns out that $\nu$ also affects, to some extent, non-uniqueness of the solution. More specifically, a large $\nu$ ensures the uniqueness of the solution, but the estimates will be biased; on the other hand, a small $\nu$ can lead to severe non-convexity causing possible local optima in the objective function being optimized. Non-uniqueness of solution is a typical problem of non-convex penalties, such as SCAD or MCP. For instance, the R package ``\textit{ncvreg}" \citep{breheny2011coordinate} warns the user when the final fit might be non-global sub-solutions. Finding the minimum value of $\nu$ which guarantees uniqueness and numerical stability is critical. It can be shown (we omit the details here) that it is possible to easily find the smallest value of $\nu$ for any value of $\lambda$ such that the solution is unique. We indicate this value as $\nu_{min}=\nu_{\lambda}$.

The penalized likelihood to be optimized at fixed $\nu$ and $\lambda$ is  

\begin{equation} \label{eq:ourpenlik}
    \ell_\lambda(\beta) = \ell(\beta) - \lambda \sum_{j=1}^p  \sqrt{2\pi} \nu \Phi\left( \frac{|\beta_j|}{\nu} \right). 
\end{equation}

$\lambda\ge 0$ is the usual tuning parameter to be selected via (generalized) CV or AIC/BIC. To estimate the coefficients, we propose to use the Alternating Direction Multiplier Method (ADMM), an algorithm that solves complex optimization problems by decomposing them into several smaller, easy-to-handle problems (for more details, refer to \cite{boyd2011distributed}).

\section{Simulation evidence}

%Lambda più piccolo in quanto le stime "esplodono".

We run some simulation experiments to compare our \textsc{CDF} penalty with respect to LASSO, MCP, and SCAD. We simulated data consisting of $n=50$ observations and $p=100$ variables from

$$ 
y_i = \beta_0 +  \sum_{j=1}^{p} x_{ih} \beta_h + \sigma_s \epsilon_i, 
$$
where $\epsilon_i\sim\mathcal{N}(0,1)$  and the covariates  $x_i \sim \mathcal{N}(0,\Sigma)$ with the Toeplitz correlation matrix $\Sigma_{jk}=0.5^{|j-k|}$. The $\sigma_s$ are four different values $(0.25, 0.5, 0.75, 1)$ of the random noise variance regulating the signal-to-noise ratio. Out of the $p=100$ parameters, only ten are non-null; for each scenario, the locations are randomly chosen, and the values of the non-null coefficients are drawn randomly from a $Unif(2, \, 2.5)$. 

Three measures are considered to compare the simulation results, namely
$$
\mathrm{MSE} = \frac{1}{B} \frac{1}{p} \sum_{j=1}^{p} \sum_{b=1}^{B} \left(\hat{\beta}_{b,j}-\beta_{j}  \right)^2 \quad \mathrm{FPR} = \frac{ \# (\hat{\beta}_{bj} \ne 0 | \beta_{j} = 0)}{  \# (\beta_{j} = 0)} \quad \mathrm{TPR} = \frac{ \# (\hat{\beta}_{bj} \ne 0 | \beta_{j} \ne 0)}{  \# (\beta_{j} \ne 0)},
$$
where $b$ is the index related to the replicates and $\hat{\beta}$ are the estimated coefficients. 
We ran 500 replicates for each scenario, and for each replicate, we fit regression models with SCAD $(\gamma=3)$ and MCP $(\gamma=3.7)$ penalties using the ``\textit{ncvreg}" package \citep{breheny2011coordinate} and the \textsc{CDF} penalty with three different values of $\nu$: $\nu=3$ (approximating the LASSO),  $\nu_{min}$, and the third value in the middle, value $\bar\nu = (\nu_{min} +3)/2$. $\bar\nu$ is used just to have a rough idea about the estimator behaviour when $\nu$ changes.

We do not discuss the selection of the best $\lambda$ values. While this point is critical in data analysis, our goal here is to compare the aforementioned penalties across the whole $\lambda$ path, i.e. from $\lambda=0$ at $\lambda_\mathrm{max}$ when all estimates are zero.

\begin{figure}[!hb]
    \centering
    \includegraphics[width=0.5\textwidth]{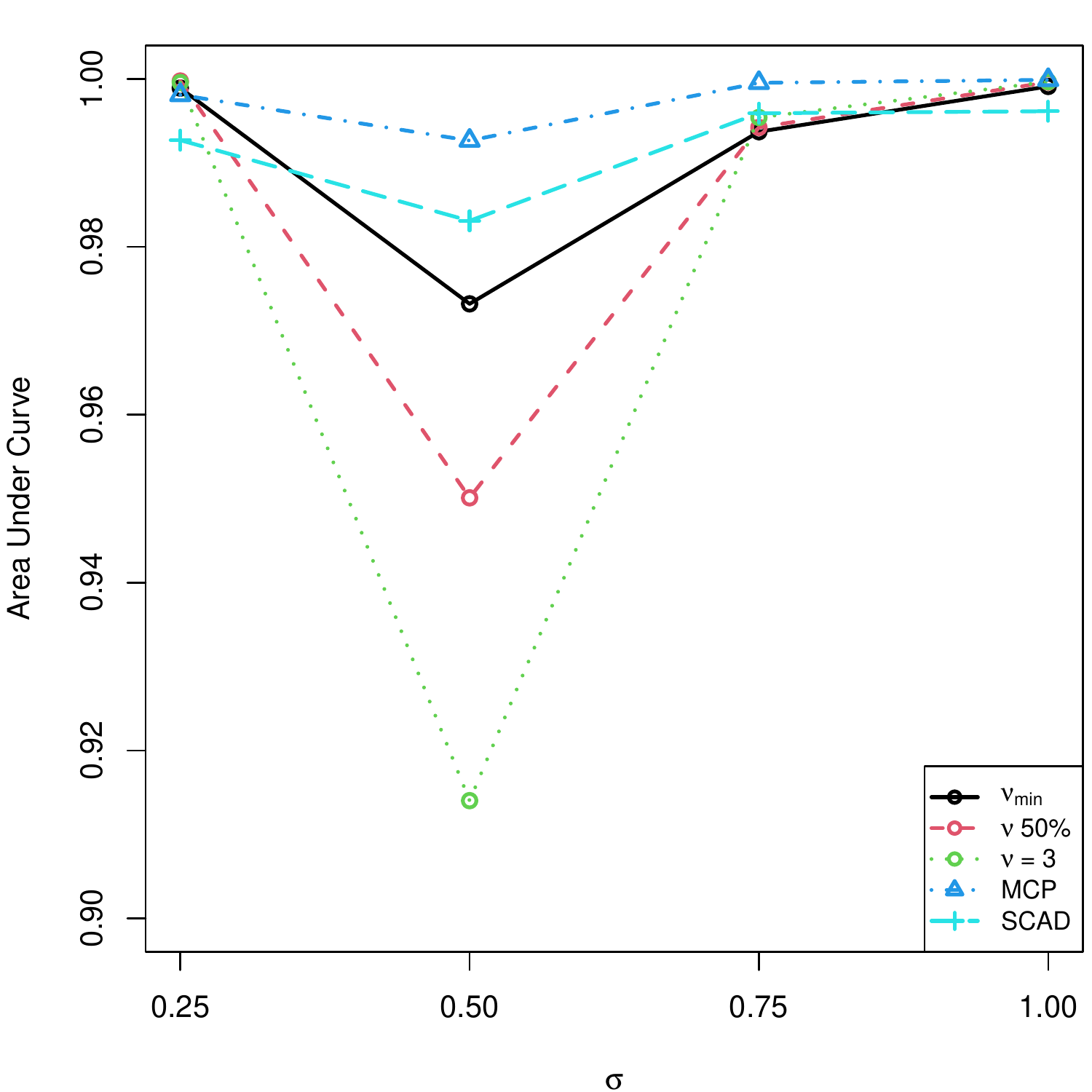}
    \caption{\label{fig:ROC} AUC curves for our proposal (considering $\nu_{min}$ and $\nu = 3$), by $\sigma$. Average over 500 replications.}
\end{figure}

Figure \ref{fig:ROC} displays the average AUC values (across 500 replicates) corresponding to the four $\sigma$ values used. The reported values of AUC are rather similar to one another, except for ones obtained at $\sigma= 0.5$, where the AUC values obtained using our proposal show a drop which is less marked when $\nu=\nu_min$. However differences were quite small, for instance the distance from the values obtained using the MCP penalty is less than 0.06. Overall, the various approaches used to select the subset of truly useful variables are substantially equivalent. 

\begin{figure}[!t]
    \centering
    \includegraphics[width=.8\textwidth]{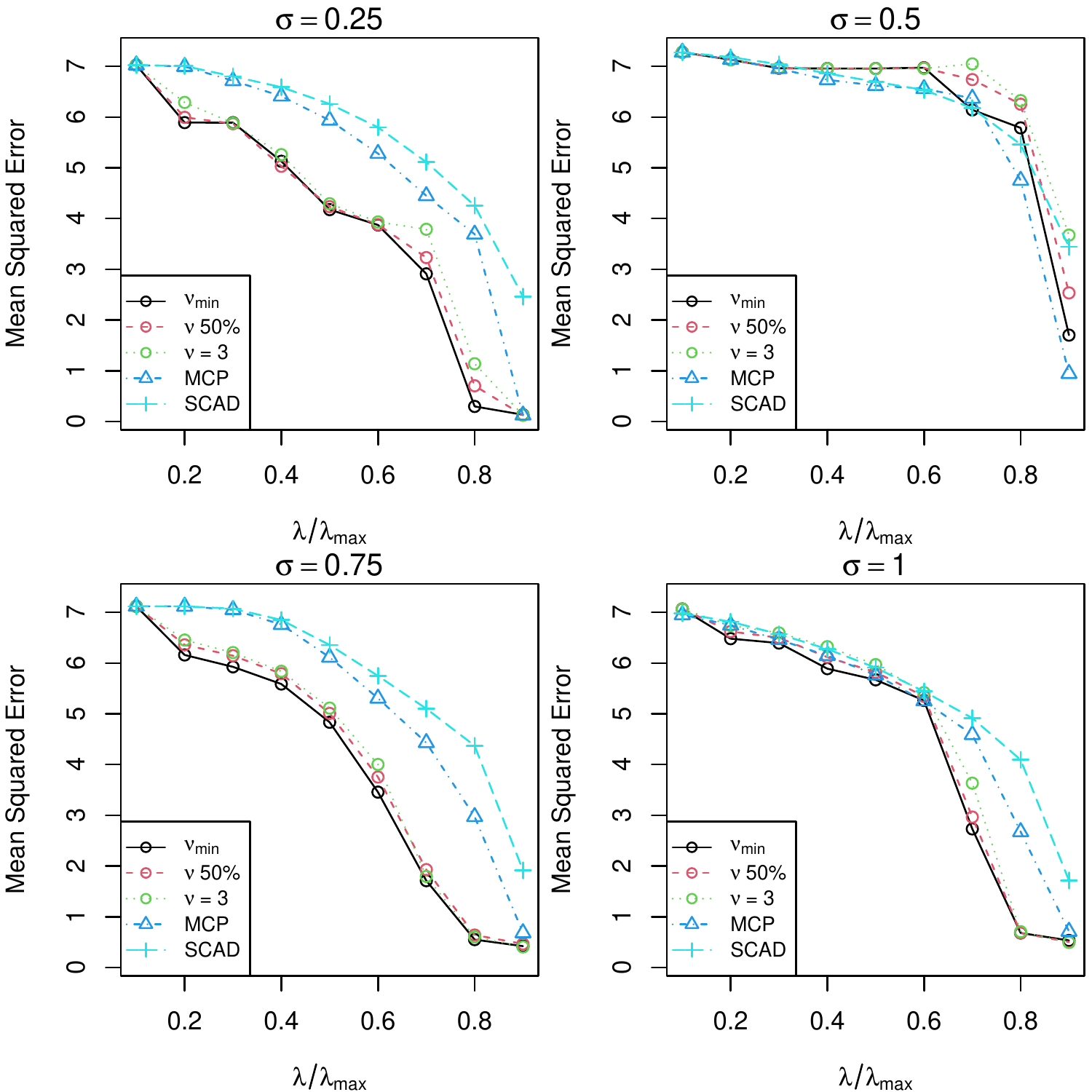}
    \caption{\label{fig:MSE} MSE curves for our proposal,  SCAD and MCP. Average over 500 replications.}
\end{figure}

%Initially, the selection capabilities of the considered models are evaluated. Figure \ref{fig:ROC} shows the ROC curve for the setting with $\beta_{min}$ equal to 1 (so the $\beta$ are generated an Unif(1, 1.5)) and $\sigma$ equal to 1. As can be guessed from the reported AUC value, our proposal with $\nu_{min}$ is the one that can best identify the subset of coefficients other than 0, while the values obtained with the highest value of $\nu$ have the worst performance. Note how the performances of the four penalized models are not far apart from each other. The reader is referred to the Supplementary Material to view the ROC curves of all hypothesized scenarios, considering additional values of $\nu$. The performance of the model estimates is then assessed in terms of Mean Squared Error. 
Figure \ref{fig:MSE} portrays the MSE along the path of $\lambda$ which were rescaled in [0,1] to make the results comparable.

As in Figure \ref{fig:ROC}, no large differences show up in the MSE values. However the proposed CDF with $\nu=\nu_{min}$ performs somewhat better in 3 out of the considered scenarios 

%This is because once the truly useful variables are entered, their estimates instantly explode to those of maximum likelihood; similarly, when a new variable is mistakenly selected for a given value of $\lambda$, it explodes in intensity too, significantly increasing the value of MSE.

%It is evident how the MSE obtained using the maximum $\nu$ is the highest (thus with estimates nearly identical to the LASSO) in the entire $\lambda$ path due to the bias component of the estimated parameters. In the left part of the graph, the best results are obtained with SCAD and MCP, but in the final part (when the $\lambda$ is closest to 0), the results obtained with our proposal using $\nu_{min}$ are the best, where the lowest overall value of the MSE is obtained. To evaluate the results in all scenarios and for different values of $\nu$, we refer the reader to the Supplementary Material.

\section{Conclusion}

In this paper, we have introduced a new penalty, we name the CDF penalty, to carry out variable selection in regression modelling through penalized likelihood. The rationale of the new penalty and relative advantages with respect to traditional LASSO, SCAD and MCP have been discussed. Simulations have shown that the \textsc{CDF} penalty performs no worse than its competitors when identifying the subset of relevant variables. 

\bibliography{main.bib}
\bibliographystyle{apalike}

\end{document}